\documentclass[aps, prb, preprint, preprintnumbers, superscriptaddress, amssymb]{revtex4-2}

\usepackage{graphicx}
\usepackage{dcolumn}
\usepackage{bm}
\usepackage{amsmath, amssymb}
\usepackage{newtxtext, newtxmath}
\usepackage{mathrsfs}
\usepackage{comment}

\def\degree{\kern-.2em\r{}\kern-.3em}

\begin{document}

\preprint{Ver. 7}

\title{CeFe$_2$Al$_{10}$: a Correlated Metal with a Fermi Surface Exhibiting Nonmetallic Conduction}


\author{Taichi Terashima}
\email{TERASHIMA.Taichi@nims.go.jp}
\affiliation{International Center for Materials Nanoarchitectonics, National Institute for Materials Science, Tsukuba, Ibaraki 305-0003, Japan}
\author{Hishiro T. Hirose}
\affiliation{Research Center for Functional Materials, National Institute for Materials Science, Tsukuba, Ibaraki 305-0003, Japan}
\author{Naoki Kikugawa}
\affiliation{Center for Green Research on Energy and Environmental Materials, National Institute for Materials Science, Tsukuba, Ibaraki 305-0003, Japan}
\author{Shinya Uji}
\affiliation{International Center for Materials Nanoarchitectonics, National Institute for Materials Science, Tsukuba, Ibaraki 305-0003, Japan}
\author{David Graf}
\affiliation{National High Magnetic Field Laboratory, Florida State University, Tallahassee, FL 32310, USA}
\author{Hitoshi Sugawara}
\email{sugawara@crystal.kobe-u.ac.jp}
\affiliation{Graduate School of Science, Kobe University, Kobe 657-8501, Japan}

\date{\today}
\begin{abstract}
Metals can be defined as materials with a Fermi surface or as materials exhibiting metallic conduction (i.e., $\mathrm{d} \rho / \mathrm{d}T > 0$).
Usually, these definitions both hold at low temperatures, such as liquid-helium temperatures, as the Fermi energy is sufficiently larger than the thermal energy.
However, they may not both hold in correlated electron systems where the Fermi energy is reduced by renormalization.
In this paper, we demonstrate that although the resistivity of CeFe$_2$Al$_{10}$ increases with decreasing temperature below $\sim20$ K, CeFe$_2$Al$_{10}$ is a metal with a Fermi surface.
This assertion is based on the observation of Shubnikov--de Haas oscillations and a Hall resistivity that changes sign with the magnetic field, which requires the coexistence of electron and hole carriers.
Our analysis of Shubnikov--de Haas and magnetotransport data indicates that the Fermi energies are as small as $\sim$30 K and that, despite the increasing carrier mobility with decreasing temperature as in conventional metals, the loss of thermally excited carriers leads to nonmetallic conduction below $\sim20$ K.
Furthermore, we investigate how this anomalous metal transforms to a more conventional metal with metallic conduction by the application of high pressure and a high magnetic field.
\end{abstract}

\maketitle
\newpage


\section{introduction}
Rare-earth and actinide compounds are fertile ground for correlated electron systems.
Their two contrasting ground states are heavy-fermion metals and Kondo insulators.
The $f$ electrons of rare-earth and actinide elements are nearly localized at high temperatures; however, they hybridize with conduction electrons at low temperatures.
Hybridization results in heavy quasiparticles at the Fermi level in the heavy-fermion metals, while the Fermi level falls in the hybridization gap in the Kondo insulators \cite{Aeppli92CCMP}.
The relevant energy scales are subject to renormalization due to electronic correlations; therefore, the Fermi energy and hybridization gap in these compounds can be much smaller than in typical metals and semiconductors.
Between these two opposite ground states is a third intriguing ground state, namely, Kondo semimetals.
In Kondo semimetals, the hybridization gap fails to fully open, leaving small Fermi pockets.
The most documented Kondo semimetal is CeNiSn.
CeNiSn was initially thought to be a Kondo insulator;
however, as the sample quality improved, the increase in resistivity at low temperatures was suppressed \cite{Nakamoto95JPSJ}.
Finally, the existence of a Fermi surface was proved by Shubnikov--de Haas (SdH) oscillations \cite{Terashima02PRB}. 
Theoretically, it was argued that the hybridization gap vanished along certain directions in the Brillouin zone \cite{Ikeda96JPSJ}.
Kondo insulators and semimetals have attracted renewed interest because they may serve as novel topological materials with strong electronic correlations \cite{Dzero10PRL, Chang17NatPhys, Lai18PNAS}.
In this study, we demonstrate that CeFe$_2$Al$_{10}$ is a Kondo semimetal.

CeFe$_2$Al$_{10}$ crystallizes in a centrosymmetric orthorhombic structure (space group $C_{mcm}$, \#63) \cite{Thiede98JMC}.
The magnetic susceptibility exhibits Curie--Weiss behavior down to 100 K with an effective moment of 2.78 $\mu_B$, approximately corresponding to the free Ce$^{3+}$ value (2.54 $\mu_B$).
The susceptibility exhibits a broad maximum at 70 K, suggesting a Kondo temperature of $T_K \simeq$ 360 K \cite{Muro09JPSJ}.
No magnetic order was detected in muon spin rotation ($\mu$SR) measurements down to 45 mK \cite{Adroja13PRB}.
The resistivity has a logarithmic temperature dependence down to 70 K, where it exhibits a broad maximum \cite{Muro09JPSJ}.
This behavior is typical of dense Kondo systems.
The resistivity then exhibits an increase below 20 K [see Fig. 2(a)]; therefore, CeFe$_2$Al$_{10}$ was regarded as a Kondo insulator \cite{Muro09JPSJ}.
An energy gap of 15 K was estimated from an Arrhenius plot between 10 and 20 K in \cite{Muro09JPSJ}.
A Schottky-like anomaly at 30 K observed in specific-heat data also supports a gap opening \cite{Muro09JPSJ}.
However, it should be noted that a finite Sommerfeld coefficient $\gamma$ of 14 mJ/(mol K$^2$) was observed as $T \rightarrow 0$ \cite{Muro09JPSJ}.
The Drude weight observed at a low temperature of 10 K in an optical study \cite{Kimura11JPSJ}, a large Knight shift \cite{Chen10PRB}, and a spin-lattice relaxation rate $T_1$ following the Korringa law below 20 K \cite{Kawamura10JPSJ} reported in nuclear quadrupole/magnetic resonance measurements further suggest that the hybridization gap is not fully open but is a pseudogap with a finite density of states at the Fermi level.
It was reported that a moderate pressure of 20 kbar caused the low-temperature resistivity to be metallic ($\mathrm{d} \rho / \mathrm{d}T > 0$) \cite{Nishioka09JPSJ}.

In this paper, we present comprehensive magnetotransport measurements on CeFe$_2$Al$_{10}$ and demonstrate that it is a Kondo semimetal, not a Kondo insulator.
We observe clear SdH oscillations, which serve as clear evidence for a Fermi surface.
We also observe that the Hall resistivity at low temperatures changes sign as the magnetic field increases.
This indicates the coexistence of electrons and holes and thus implies an intrinsic semimetallic electronic structure with overlapping electron and hole bands.
Based on SdH data and multicarrier analysis of resistivity and Hall resistivity data, we demonstrate that the Fermi energies are as small as $\sim$ 30 K and ascribe the nonmetallic behavior ($\mathrm{d} \rho / \mathrm{d}T < 0$) below $\sim$20 K to the loss of thermally excited carriers.
We further demonstrate that there is no metal-insulator phase transition separating the nonmetallic-conduction state and a conventional metallic-conduction state by performing high-pressure and high-magnetic-field measurements.
We show that nonmetallic conduction smoothly transforms into metallic conduction without phase transition as high pressures or high magnetic fields are applied.

\section{Experimental methods}
\subsection{Sample preparation}
Single crystals of CeFe$_2$Al$_{10}$ were grown by an Al self-flux method as described in \cite{Takesaka10JPCS}.
The starting materials were 3N (99.9\% pure) Ce, 4N Fe, and 5N Al.
We employed a starting molar ratio of Ce:Fe:Al = 1:2:20$\sim$30. 
Fig. 1(a) shows a photograph of a grown single crystal.
The phase purity of grown crystals was confirmed by a powder X-ray pattern of crashed crystals [Fig. 1(b)].
The lattice parameters determined from single-crystal X-ray diffraction were $a$ = 9.0090(3) \AA, $b$ = 10.2373(7) \AA, and $c$ = 9.0731(8) \AA, which are consistent with a previous report \cite{Muro13jkps}.
The crystallographic axes of the samples used in this study were determined by using a back-reflection Laue method [Fig. 1(c)-(e)].

\subsection{Magnetotransport measurements}
The electrical contacts were spot-welded and reinforced with silver conducting paint.
We used a 20-T superconducting magnet equipped with a dilution refrigerator with a base temperature of 0.03 K and a 17-T superconducting magnet with a $^4$He variable temperature insert at the National Institute for Materials Science, Tsukuba, Japan, and a 45-T hybrid magnet with a $^3$He insert at the National High Magnetic Field Laboratory in Tallahassee, Florida, USA.
To exclude crystals with Al inclusions, we measured the resistivity versus temperature curves down to 0.03 K at the smallest possible field and the resistivity versus field curves around zero field at 0.03 K for all samples.
Since the superconducting magnet had a residual field of approximately 0.03 T, we carefully reduced the field at the sample position to less than 0.001 T (10\% of $B_{c2}$ of Al) by applying a small current to the magnet for the resistivity versus temperature measurements.
In addition, we measured commercial Al foil together with the samples to ensure that the superconducting transition of Al could be detected with the used setup.
All the results presented in this paper were obtained using samples that exhibited no trace of a superconducting transition and thus were Al-inclusion free. 
The used samples are listed in Table I.
The sample names (a1, a3, etc.) are composed of the longest direction along which the electrical current was applied and the sample number.
For the simultaneous resistivity and Hall resistivity measurements presented in Figs. 3(a)--(d) and 5(a)--(b), measurements were performed in both positive and negative fields, and experimental signals $\rho^{\mathrm{exp}}$ and $\rho_{H}^{\mathrm{exp}}$ were symmetrized and antisymmetrized to obtain true $\rho$ and $\rho_{H}$, respectively; that is, $\rho(B) = (\rho^{\mathrm{exp}}(B) + \rho^{\mathrm{exp}}(-B))/2$ and $\rho_{H}(B) = (\rho_{H}^{\mathrm{exp}}(B) - \rho_{H}^{\mathrm{exp}}(-B))/2$.
For the SdH data presented in Figs. 2(b) and 4(b) and the high-field data presented in Figs. 6(a) and (b), measurements were performed in positive fields.
 
\subsection{High-pressure generation}
We used a piston--cylinder-type pressure cell made of NiCrAl alloy.
The pressure-transmitting medium was Daphne 7474 (Idemitsu Kosan), which remains liquid up to 37 kbar at room temperature and ensures highly hydrostatic pressure generation in the investigated pressure range \cite{Murata08RSI}.
The pressure was determined from the resistance variation of the calibrated manganin wires.

\section{Results}
\subsection{Shubnikov--de Haas oscillation at ambient pressure}
Figure 2(a) presents the temperature dependence of the $a$-, $b$-, and $c$-axis resistivities measured on samples a1, b4, and c3.
Consistent with a previous single-crystal study \cite{Tanida14JPSJ}, the resistivity increased below $\sim$20 K for all three axes.
The increase in resistivity continued down to 0.04 K, the lowest measurement temperature.
Nevertheless, we observed clear SdH oscillations.
Figure 2(b) presents the magnetoresistance for two field orientations $B \parallel b$ and $B \parallel c$ measured on sample a3 at $T$ = 0.04 K.
The positive magnetoresistance indicates that the magnetoresistance was dominated by the orbital motion of carriers.
SdH oscillations were already visible in the raw data before background subtraction.
The upper left inset displays the oscillatory part $\Delta\rho$ normalized by a smooth background $\rho_0$, which is given by a third-order polynomial.
The lower right inset displays the Fourier transform of $\Delta\rho / \rho_0$ versus $1/B$ in the field range 5--17.5 T.
Clear peaks in the Fourier spectra indicate that the oscillations were periodic in $1/B$ in this field range and therefore that the magnetic fields in this range and at these orientations did not alter the electronic band structure except for trivial Zeeman energy shifts.

For $B \parallel b$, two frequency peaks appeared at $F$ = 28(4) and 54(4) T.
The SdH frequency $F$ is related to the Fermi-surface cross-sectional area $A$ as $F=(\hbar/2\pi e)A$. 
It should be noted that the higher frequency is twice the lower frequency within the experimental accuracy.
Figures 2(c) and (d) plot the temperature dependences of the peak amplitudes of the two frequencies.
According to the Lifshitz--Kosevich formula, the temperature dependence is described by $R_{T, r}= rX/\sinh(rX)$, where $X=K(m^*/m_e) T/B$, $K=2\pi^2k_Bm_e/(e\hbar)$ (= 14.69 T/K), and $r$ is the harmonic number \cite{Shoenberg84}.
We temporarily assumed $r$ = 1 for both frequencies, fit $R_{T, r}$ to these dependences (black lines), and obtained the effective masses of $m^*$ = 0.91(9)$m_e$ and 2.1(2)$m_e$ for the two frequencies.
The latter is twice the former within the experimental accuracy.
This confirms that the higher frequency is the second harmonic ($r$ = 2) of the lower frequency.
By combining these two effective masses, the most probable estimate is $m^*$ = 1.01(5)$m_e$ for $B \parallel b$ (sample a3).
Figure 2(e) presents the magnetic-field dependence of the second-harmonic amplitude as a Dingle plot for $B \parallel b$.
The magnetic-field dependence is given by $R_{D,r}=\exp(-rX_D)$, where $X_{D}=K(m^*/m_e) T_{D}/B$ \cite{Shoenberg84}.
The Dingle temperature $T_D$ is defined as $T_D=\hbar/(2\pi k_B\tau)$, where $\tau$ is the scattering time; therefore, $X_D$ can be written as $X_{D}=\pi/(|\mu|B)$ using the mobility $\mu$.
The solid line is a fit of this Dingle reduction factor to the experimental data points. From its slope, we obtained a Dingle temperature $T_D$ of 1.01(6) K ($\tau$ = 1.20$\times 10^{-12}$ s) or a mobility $|\mu|$ of 2.12(7) $\times10^3$ cm$^2$V$^{-1}$s$^{-1}$ (sample a3).

For $B \parallel c$, only one frequency was observed at $F$ = 46(4) T [Fig. 2(b), lower right inset].
From a similar analysis, we obtained $m^*$ = 1.87(6)$m_e$, $T_D$ = 0.98(5) K, and $|\mu|$ = 2.2(1) $\times10^3$ cm$^2$V$^{-1}$s$^{-1}$ (sample a3).

Figure 2(f) presents the magnetic-field-angle dependence of the Fourier spectra, where the angle $\theta$ was measured from $B \parallel b$.
The SdH measurements were performed every 5$^{\circ}$ of $\theta$.
The white circles indicate the frequency peaks of the Fourier spectra.
The black line is a fit to $F(\theta) = F(0)/(a^2\sin^2\theta + \cos^2\theta)^{1/2}$, which holds for an ellipsoidal Fermi pocket, while the broken line represents the expected second-harmonic frequency.
From this fit, the Fermi pocket is estimated to occupy 2.2$\times10^{-4}$ or 2.8$\times10^{-4}$ of the Brillouin zone volume depending on whether the pocket is oblate (short axis $\parallel c$) or prolate (long axis $\parallel b$), which yields 4.3$\times10^{-4}$ or 5.5$\times10^{-4}$ carriers per pocket.
These values correspond to a carrier density of 1.0--1.3$\times10^{18}$ cm$^{-3}$ (sample a3) if we assume a single pocket in the Brillouin zone.
It is instructive to estimate the Sommerfeld coefficient $\gamma$ associated with this carrier density.
For simplicity, we assumed $m^*=m_e$; then, we obtained $\gamma$ = 0.2 mJ/(mol K$^2$), which is two orders of magnitudes smaller than the reported value of 14 mJ/(mol K$^2$) \cite{Muro09JPSJ}.

We estimated the Fermi energy of this pocket by assuming a quadratic dispersion $E_F=\hbar^2k_F^2/(2 m^*)$ and $A=\pi k_F^2$.
Using the $F$ and $m^*$ values for $B \parallel b$ [$B \parallel c$], we obtained $E_F$ = 37(7) K [33(4) K] (sample a3).

For sample a1, smaller SdH frequencies were observed; the fundamental frequency and associated effective mass were $F$ = 12(3) T and $m^*$ = 1.4 (2) $m_e$, respectively, for $B \parallel b$, and 23(5) T and 2.5(3) $m_e$, respectively, for $B \parallel c$.
The carrier density was estimated as 0.3--0.5$\times10^{18}$ cm$^{-3}$.
The Fermi energy was $E_F$ = 12(5) K irrespective of $B \parallel b$ or $c$.
Considering the very small Fermi energies and carrier densities of samples a3 and a1, their differences are likely ascribable to either slight differences in their chemical compositions or impurity and defect concentrations.

\subsection{Magnetoresistivity and Hall resistivity at ambient pressure} 
In this section, we analyze the magnetoresistivity and Hall resistivity.
Figures 3(a) and (b) present the magnetic-field dependence of the resistivity $\rho$ and Hall resistivity $\rho_H$ at $T$ = 0.04 K of sample a3 (brown solid lines).
The current was applied along the $a$ axis, while the magnetic field was applied along the $b$ axis.
The Hall resistivity was positive at low fields but became negative at high fields.
This nonmonotonic behavior indicates that there are multiple carriers and that the major carrier is an electron with low mobility that dominates at high fields.

Before proceeding with the analysis, we consider a possible anomalous Hall effect contribution from skew scattering, which can be significant in Ce compounds.
According to \cite{Fert87PRB}, the Hall constant from skew scattering is given by $R_H^{skew} = \gamma \tilde{\chi} \rho_{mag}$,
where $\gamma$ is typically $\sim$0.08 KT$^{-1}$ for Ce compounds \cite{Fert87PRB}.
The reduced magnetic susceptibility is given by $\tilde{\chi} = \chi/C$, where $C$ is the Curie constant.
Using a value of $\chi \approx 1 \times 10^{-3}$ emu/mol for $B \parallel b$ at low temperatures \cite{Tanida14JPSJ} and the Curie constant for a free Ce ion, we obtained $\tilde{\chi} \approx 1 \times 10^{-3}$ K$^{-1}$.
To estimate the upper bound on $R_H^{skew}$, we used the total resistivity $\rho \approx 2$ m$\Omega$cm at $B$ = 10 T [Fig. 3(a)] instead of $\rho_{mag}$.
Then, we obtained $R_H^{skew} \approx 0.2$ $\mu\Omega$cmT$^{-1}$, corresponding to $\rho_H \approx 0.002$ m$\Omega$cm at $B$ = 10 T.
This value is negligible compared to the magnitudes of the experimental Hall resistivities (Figs. 3 and 5).

For quantitative analysis, we used a simple multicarrier model.
We assumed that the density, charge, and mobility of the $i$-th carrier were given by $n_i$, $q_i$, and $\mu_i$, respectively.
In our formalism, $q_i$, and $\mu_i$ have a sign (+ for holes and - electrons).
We then assumed that the conductivity of the $i$-th carrier in the $xy$ plane normal to the applied field was described by the conductivity tensor
\begin{equation}
\hat{\sigma}^i= \frac{n_iq_i\mu_i}{1+(\mu_iB)^2}
\left(\begin{array}{cc}
1 & \mu_iB \\
-\mu_iB & 1 \\
\end{array} \right),
\end{equation}
where we took $x$ parallel to the current.
We calculated the total conductivity tensor $\hat{\sigma}=\sum_i \hat{\sigma}^i$ and then the resistivity tensor $\hat{\rho}=\hat{\sigma}^{-1}$.
We fit $\rho(B)$ and $\rho_H(B)$ simultaneously, using $n_i$, $q_i$, and $\mu_i$ as fitting parameters.

We first attempted to use a two-carrier model ($i$ = 1 and 2) but found that it could not explain the experimental data.
The dash-dotted lines in Figs. 3(a) and (b) display the least-squares fitting results of the two-carrier model.
We therefore used a three-carrier model ($i$ = 1, 2, and 3), which provided excellent results.
The gray dashed lines in Figs. 3(a) and (b) display the three-carrier fit to the data, which almost completely overlap the brown experimental lines.
The fits indicate the existence of minor hole (h) and electron carriers (e1) and a major electron carrier (e2).
For the minor hole and electron, [$n_i$ ($10^{18}$ cm$^{-3}$), $\mu_i$ (cm$^2$V$^{-1}$s$^{-1}$)] = [1.162(3), 1690(3)] and [1.196(4), --1336(4)], respectively, for sample a3 (the errors are numerical fitting errors).
These carrier densities and mobilities are in reasonable agreement with the values deduced from the SdH data.
This indicates that the observed SdH frequency was due to either the minor hole or electron.
It is unclear why the other minor carrier was not observed in the SdH oscillations; however,
this may have been due to an undesirable Fermi pocket shape or spin splitting \cite{Shoenberg84}.
Alternatively, the frequencies from the two minor carriers may have been too close to be resolved (i.e., the observed SdH oscillations are a superposition of the two close frequencies).
For the major electron, [$n_i$ ($10^{18}$ cm$^{-3}$), $\mu_i$ (cm$^2$V$^{-1}$s$^{-1}$)] = [53.3(7), --47.8(7)].
The carrier density corresponds to 0.0110(2) electrons per formula unit.
Because of the much smaller mobility of the major electron carrier than that of the minor carriers, it is reasonable that SdH oscillations due to this carrier were not observed.
However, the existence of a major carrier with a low mobility is supported by the reported Sommerfeld coefficient of 14 mJ/(mol K$^2$). 
Let us assume a spherical Fermi pocket and $m^*=20 m_e$ for this carrier, as the mobility was more than 20 times smaller than that of the minor carriers.
Then, we obtained a Fermi energy of 30 K and a Sommerfeld coefficient of 15 mJ/(mol K$^2$) for this carrier (sample a3).

Since sample a3 was measured using a dilution refrigerator, high-temperature measurements were impossible.
We therefore measured another sample, sample a11, from $T$ = 1.4 to 120 K in a $^4$He variable-temperature insert [Figs. 3(c) and (d)].
The current was applied along the $a$ axis, while the magnetic field was applied along the $b$ axis.
The colored solid curves in Figs. 3(c) and (d) represent the experimental data, while the dashed curves represent the three-carrier fitting results, which almost completely overlap the former.
Although the Hall resistivity remained positive in the measured field range for $T \geqslant 40$ K, the use of the three-carrier model was justified because the fit parameters evolved smoothly up to 120 K [Figs. 3(e)--(g)].
By switching to a two- (or one)-carrier model, unphysical jumps will occur in the parameters.
Furthermore, even at $T$ = 120 K, the three-carrier model produces a 20\% smaller chi-square value than that produced by the two-carrier model.
The fitted curves for $T \geqslant 40$ K, as extrapolated, turned negative at fields higher than the measured field range.

At $T$ = 1.4 K, [$n_i$ ($10^{18}$ cm$^{-3}$), $\mu_i$ (cm$^2$V$^{-1}$s$^{-1}$)] = [0.359(1), 1213(2)], [0.275(1), -1176(3)], and [58.4(3), -44.6(2)] for the three carriers (sample a11).
The carrier densities of the minority carriers were in reasonable agreement with those estimated from the SdH oscillations of sample a1.

Figures 3(e)--(g) present the temperature variation of the density, mobility, and zero-field conductivity $\sigma_i=n_i q_i \mu_i$ for the three carriers in sample a11.
Although the resistivity increased with decreasing temperature below $\sim$20 K, the magnitude of the mobility $|\mu|$ increased with decreasing temperature down to 4 K for the minor hole and electron carriers and down to 10 K for the major-electron carrier [Fig. 3(f)].
The mobilities were approximately constant below those temperatures.
However, the carrier density decreased for all three carriers as the temperature was lowered [Fig. 3(e)], leading to a decrease in the conductivity of each carrier below $T \sim$20 K [Fig. 3(g)].
Therefore, the resistivity increase below $\sim$20 K is ascribed to the rapid decrease in the carrier density [Fig. 3(e)].

The temperature dependence of the major electron (e2) density can be fitted to activation-type behavior $n=n_0 + n_1 \exp(-E_{ex}/k_B T)$.
The solid line in Fig. 3(e) is a fit to the data up to $T$ = 20 K, while the dashed line is an extrapolation to higher temperatures.
The obtained excitation energy is $E_{ex}$ = 28(1) K (the error is a numerical fitting error).

\subsection{Pressure dependence}
In this section, we report SdH and magnetotransport measurements under high pressure, which show that the nonmetallic conduction smoothly transforms into a conventional metallic conduction without a phase transition as the carrier number continuously increases with pressure.

We first examine the pressure evolution of SdH oscillations.
Figure 4(a) presents the temperature dependence of the resistivity of sample a3 ($I \parallel a$) measured at various pressures up to $P$ = 20.3 kbar.
The $\rho$ versus $T$ curves were recorded while a top-loading probe was inserted into a dilution refrigerator; thus, the temperature could not be measured accurately due to the temperature difference between the sample and thermometer.
Accordingly, the curves exhibit many kinks.
Nevertheless, it is clear that the low-temperature resistivity increase was suppressed as the pressure was increased.
The resistivity was metallic (d$\rho$/d$T$ > 0) at low temperatures at $P$ = 9.4 kbar (light blue) and above, consistent with a previous report \cite{Nishioka09JPSJ}.

Figure 4(b) presents the SdH oscillations for $B \parallel b$ at various pressures measured on sample a3.
The 0-kbar data are the same as the $B \parallel b$ curve in the upper inset of Fig. 2(b) but are plotted as a function of 1/$B$.
The corresponding Fourier spectra are provided in Fig. 4(c).
In general, as the pressure increased, the oscillations became faster [Fig. 4(b)], and major frequency components increased [Fig. 4(c)].
The pressure of 2.7 kbar appeared to be an exception: the oscillation [orange curve in Fig. 4(b)] appeared slower than that at $P$ = 0 kbar.
However, there was a frequency peak at $F$ = 84 T in the spectrum [Fig. 4(c)], which was higher than the two frequencies at $P$ = 0 kbar.
Therefore, the trend of increased carrier density with increased pressure was likely observed.
Figure 4(d) plots $F^{3/2}$ as a measure of the carrier density for frequency peaks observed at each pressure.
The solid and open circles correspond in Figs. 4(c) and (d).
We determined the effective mass for most of the observed frequencies.
The masses ranged from $\sim$1 to 2 $m_e$, except for 3.3 $m_e$ for the $F$ = 211(8) T peak at $P$ = 5.8 kbar, which was most likely a second harmonic of the $F$ = 104(8) T peak.
No clear increasing or decreasing trend with pressure was observed.

We now analyze the pressure variation of the magnetotransport data.
Figures 5(a) and (b) present the transverse magnetoresistivity and Hall resistivity (colored solid curves) measured on sample a3 with $B \parallel b$ at $T$ = 0.03 or 0.04 K except for $P$ = 16.7 kbar, where measurements were performed at $T$ = 0.22 K.
We used the same three-carrier model as above, and the fitting results are indicated by dashed gray curves, which almost completely overlap the experimental curves.
Figures 5(c)--(e) display the carrier density, mobility, and conductivity obtained from the fits. 
The pressure of 2.7 kbar was again somewhat peculiar: the carrier density of the hole (red) had a local minimum, while that of the major electron (electron 2, green) had a local maximum.
Except for these carrier densities, the carrier densities generally increased with pressure.
The magnitude of the hole mobility (red) decreased with pressure, while that of the major-electron mobility (green) increased.
That of the minor-electron mobility (blue) was approximately constant.
Despite these variations in mobility, all the conductivities increased with pressure [Fig. 5(e)] except for the hole conductivity (red), which had a local minimum at 2.7 kbar.

Figure 4(d) compares the carrier densities of the minor hole and electron with $F^{3/2}$ from the SdH data.
The left and right vertical scales were adjusted so that the black $F^{3/2}$ point and blue electron density overlap at $P$ = 0 kbar.
The figure demonstrates that the increase in carrier density with pressure inferred from the SdH data and that observed from the magnetotransport data are in reasonable agreement with respect to the minority carriers.
This correspondence confirms that the observed SdH oscillations are intrinsic to CeFe$_2$Al$_{10}$:
if the SdH oscillations were from impurities or inclusions in samples, no such correspondence would be observed.

\subsection{Magnetic-field-induced metallization}
Lastly, we present the magnetoresistivity data up to a magnetic field of 45 T, which again shows a smooth transformation from the nonmetallic-conduction state to a metallic-conduction state with increasing field.
Figures 6(a) and (b) present the transverse magnetoresistivity measured on samples a1 and b4, respectively.
Since measurements up to 45 T were performed with a hybrid magnet, low-field data below 11.5 T were lacking.
We therefore added resistivity data up to 17.5 T measured at $T$ = 0.03 K in a dilution refrigerator.
The low-field behavior was generally consistent with a previous report \cite{Tanida14JPSJ}.
For sample a1 ($I \parallel a$), the resistivity started decreasing above $\sim$37 T for $B \parallel b$ (dashed curve) and above $\sim$21 T for $B \parallel c$ (solid curve) at $T$ = 0.4 K [Fig. 6(a)].
This indicates that magnetic fields of these strengths in these directions caused nontrivial changes in the electronic band structure, such as a new band crossing the Fermi level.
The temperature dependence of the resistivity remained nonmetallic even at $B$ = 45 T applied along the $c$ axis.
For sample b4 ($I \parallel b$), the resistivity started decreasing above $\sim$11 T for $B \parallel c$ (dashed curve) at $T$ = 0.03 K [Fig. 6(b)].
The resistivity at $T$ = 0.4 K exhibited a fairly large decrease as the field was increased to 45 T; however, the temperature dependence remained nonmetallic.
As the field was applied along the $a$ axis, the resistivity at $T$ = 0.03 K (solid curve) started decreasing at $\sim$1 T and exhibited a pronounced decrease with increasing field.
Furthermore, the temperature dependence was inverted at $B \sim$27 T; namely,
the apparently nonmetallic state (d$\rho$/d$T$ < 0) at low fields transformed into a conventional metallic state (d$\rho$/d$T$ > 0) at high fields without phase transition.

In the order of $B \parallel b$, $c$, and $a$, the onset field of negative magnetoresistance decreased, and the magnitude of the negative magnetoresistance increased.
This correlates with the anisotropy of magnetic susceptibility, which increases in the same order \cite{Tanida14JPSJ}.


\section{Discussion}
The quantum oscillation originates from Landau quantization of the orbital motion of charged carriers in magnetic fields \cite{Landau30ZPhys}.
In accord with this, our analyses of the SdH oscillations and magnetotransport data indicated that the observed oscillations are due to either the minor hole (h) or electron (e1) carriers.

Previously, it was reported that a Kondo insulator SmB$_6$ exhibited quantum oscillations \cite{Li14Science, Tan15Science}.
However, in those reports the connection between the observed oscillations and charged carriers is unclear, and the origin of the oscillations is highly debated \cite{Thomas19PRL}.

Some theories suggested that the quantum oscillation could be observed in insulators without Fermi surface \cite{Knolle15PRL, Zhang16PRL}.
Those theories argued that the temperature dependence of the oscillation amplitude would deviate from the Lifshitz--Kosevich formula describing the quantum oscillation in metals with Fermi surface:
Ref.~\cite{Knolle15PRL} suggested that the amplitude might have a maximum at a temperature set by a hybridization gap or that it might exhibit a characteristic steep increase at lowest temperatures, while Ref.~\cite{Zhang16PRL} suggested that the temperature dependence of the amplitude might have an unusual two-plateau feature.
However, we saw no such features in our data.
The temperature dependences of the oscillation amplitudes observed in the present study nicely follow the Lifshitz--Kosevich formula as exemplified in Figs. 2(c) and (d).
Furthermore, the fact that the effective mass of the second harmonic is twice that of the fundamental and also the satisfactory Dingle plot shown in Fig. 2(e) conform to the Lifshitz--Kosevich formula and corroborate that the observed Shubnikov--de Haas oscillations are due to charged carriers on the Fermi surface.

Figure 7 presents a schematic of the electronic structure near the Fermi level deduced from the experimental results, displaying light electron (e1) and hole (h) bands.
The Fermi energy of either band $E_F^{e1}$ or $E_F^{h}$, was estimated to be 33(4)--37(7) K for sample a3 and 12(5) K for sample a1.
The Fermi energy of the other band can be assumed to be of similar size.
In addition, there was a heavy electron band (e2) whose Fermi energy $E_F^{e2}$ was estimated to be 30 K (32 K) for sample a3 (sample a11) when $m^* = 20 m_e$ was assumed.
The activation-type behavior of the carrier concentration indicates the existence of states with a large density of states at an energy $E_{ex}$ = 28(1) K (sample a11) below the Fermi level.
These states may be a hole band with an effective mass even larger than that of the heavy electron band or incoherent states, as they do not appear in the magnetotransport data.

We note that the Zeeman energy of an electron spin corresponds to 20 K at $B$ = 30 T, which is close to the energy scales mentioned above.
Therefore, it is reasonable to assume that the magnetic-field-induced metallization is caused by the Zeeman energy.

The present results indicate that CeFe$_2$Al$_{10}$ is an intrinsic metal with overlapping electron and hole bands.
However, as the Fermi energies are small, the carrier densities are not constant but vary with temperature even in $^4$He or lower temperature ranges.
The apparent nonmetallic conduction observed below 20 K at ambient pressure is ascribed to the rapid decrease in carrier density with decreasing temperature.
As the band overlap and thus the carrier densities are increased by the application of high pressure or a high magnetic field (via Zeeman energy in the latter case), nonmetallic conduction smoothly transforms into conventional metallic conduction without phase transition.

The conclusion that CeFe$_2$Al$_{10}$ is an intrinsic metal is consistent with a previous optical study \cite{Kimura11JPSJ}, which found that although the effective carrier density decreased with decreasing temperature, a finite Drude weight still existed at $T$ = 10 K.
This conclusion is also consistent with nuclear quadrupole/magnetic resonance measurements reporting a pseudogap instead of a full gap \cite{Chen10PRB, Kawamura10JPSJ}.

Usual density-functional-theory band calculations yield large Fermi surface in CeFe$_2$Al$_{10}$ \cite{Nam21PRB}, which is inconsistent with the transport data.
A recent dynamical-mean-field-theory (DMFT) study showed that the Ce-4$f$ and conduction electron states hybridized as temperature was lowered \cite{Nam21PRB}.
Although the Fermi volume was reduced, the hybridization gap did not open fully, and CeFe$_2$Al$_{10}$ remained a metal with several bands crossing the Fermi level at $T$ = 150 K (the lowest temperature of the calculations) \cite{Nam21PRB}.
The metallic ground state is consistent with our conclusion, but the DMFT Fermi surface is still far larger than the observed tiny Fermi pocket, whose volume is only $\sim10^{-4}$ of the Brillouin zone.
The temperature of 150 K may not be sufficiently low for direct comparison with the present results, as the resistivity maximum indicating the onset of coherence is located at $\sim$70 K for polycrystals \cite{Muro09JPSJ} and $\sim$40 K for a single crystal with $I \parallel a$ [Fig. 2(a)]. 
Calculating and comparing electronic structures at lower temperatures with the present experimental results would be of interest:
it would clarify whether the ground-state electronic structure is topologically trivial or not.

The size of the pseudogap has been estimated in various studies.
For example, an optical study \cite{Kimura11JPSJ} reported a gap size of 55 meV, while nuclear quadrupole/magnetic resonance studies reported a gap size of 110 and 70 K \cite{Chen10PRB, Kawamura10JPSJ}. A photoemission spectroscopy study reported a gap size of 60 and 12 meV \cite{Ishiga14JPSJ}.
In addition, a spin gap of 12.5 meV was observed in inelastic neutron scattering measurements \cite{Adroja13PRB}.
It is unclear how these gap sizes are related to the presently determined electronic structure.

Finally, we comment on recent studies of Ce$_3$Bi$_4$Pd$_3$.
It was postulated that Ce$_3$Bi$_4$Pd$_3$ is a Weyl--Kondo semimetal based on specific-heat data \cite{Dzsaber17PRL}.
A dynamical mean-field theory study also proposed that it is a topological nodal-line semimetal \cite{Cao20PRL}.
In contrast, Ce$_3$Bi$_4$Pd$_3$ was assumed to be a Kondo insulator in a magnetotransport study \cite{Kushwaha19NatCommun}.
The authors of \cite{Kushwaha19NatCommun} observed field-induced metallization similar to what we observed in CeFe$_2$Al$_{10}$ for $B \parallel a$ [Fig. 6(b)] and argued that it might be associated with quantum criticality.
In addition, they observed an activation-type temperature dependence of the carrier density similar to what we observed in CeFe$_2$Al$_{10}$ [Fig. 3(e)].
The present investigation of CeFe$_2$Al$_{10}$ suggests that more thorough studies are necessary to determine whether Ce$_3$Bi$_4$Pd$_3$ is a topological semimetal or a Kondo insulator.

In summary, we observed clear SdH oscillations and a change in sign of the Hall resistivity with the magnetic field in CeFe$_2$Al$_{10}$ at ambient pressure.
The former is evidence of a Fermi surface, while the latter indicates an intrinsic semimetallic electronic structure with overlapping electron and hole bands.
We therefore concluded that CeFe$_2$Al$_{10}$ is a Kondo semimetal and not an insulator as previously thought.
Analysis of the resistivity and Hall resistivity indicated that the Fermi energies were as small as $\sim$30 K, and we ascribed the apparently nonmetallic conduction behavior to the loss of thermally excited carriers.
We also demonstrated that the nonmetallic behavior at low temperatures smoothly transformed into metallic behavior without phase transition as high pressures or high magnetic fields were applied.
Finally, the topological character of the ground-state electronic structure remains to be examined in future work.

\begin{acknowledgments}
This work was supported in Japan by JSPS KAKENHI (No. 22H04485, 22K03537, 18K04715, 21H01033, 22H01173, 21K03470, and 22K19093).
A portion of this work was performed at the National High Magnetic Field Laboratory, which is supported by the National Science Foundation Cooperative Agreement No. DMR-1644779 and the State of Florida.
\end{acknowledgments}

%

\newpage

\begin{figure*}
\includegraphics[width=11cm]{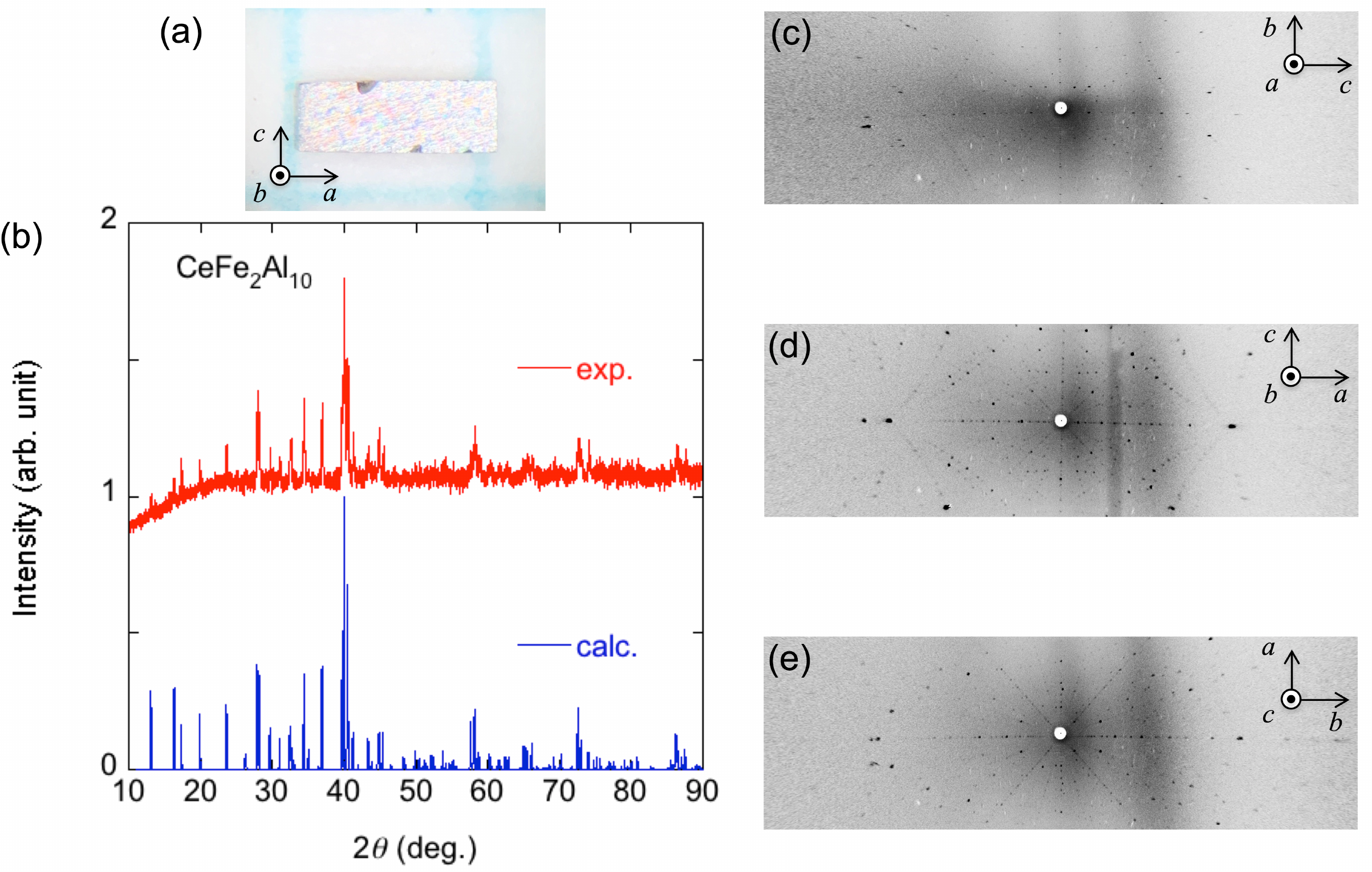}
\caption{\label{Fig1}
Al-flux grown CeFe$_2$Al$_{10}$.
(a) Photograph of sample a3.
(b) Powder X-ray diffraction pattern of crashed crystals (exp.) compared to simulation (calc.).
(c), (d), and (e) X-ray Laue patterns for sample a3.
}
\end{figure*}

\begin{figure*}
\includegraphics[width=16cm]{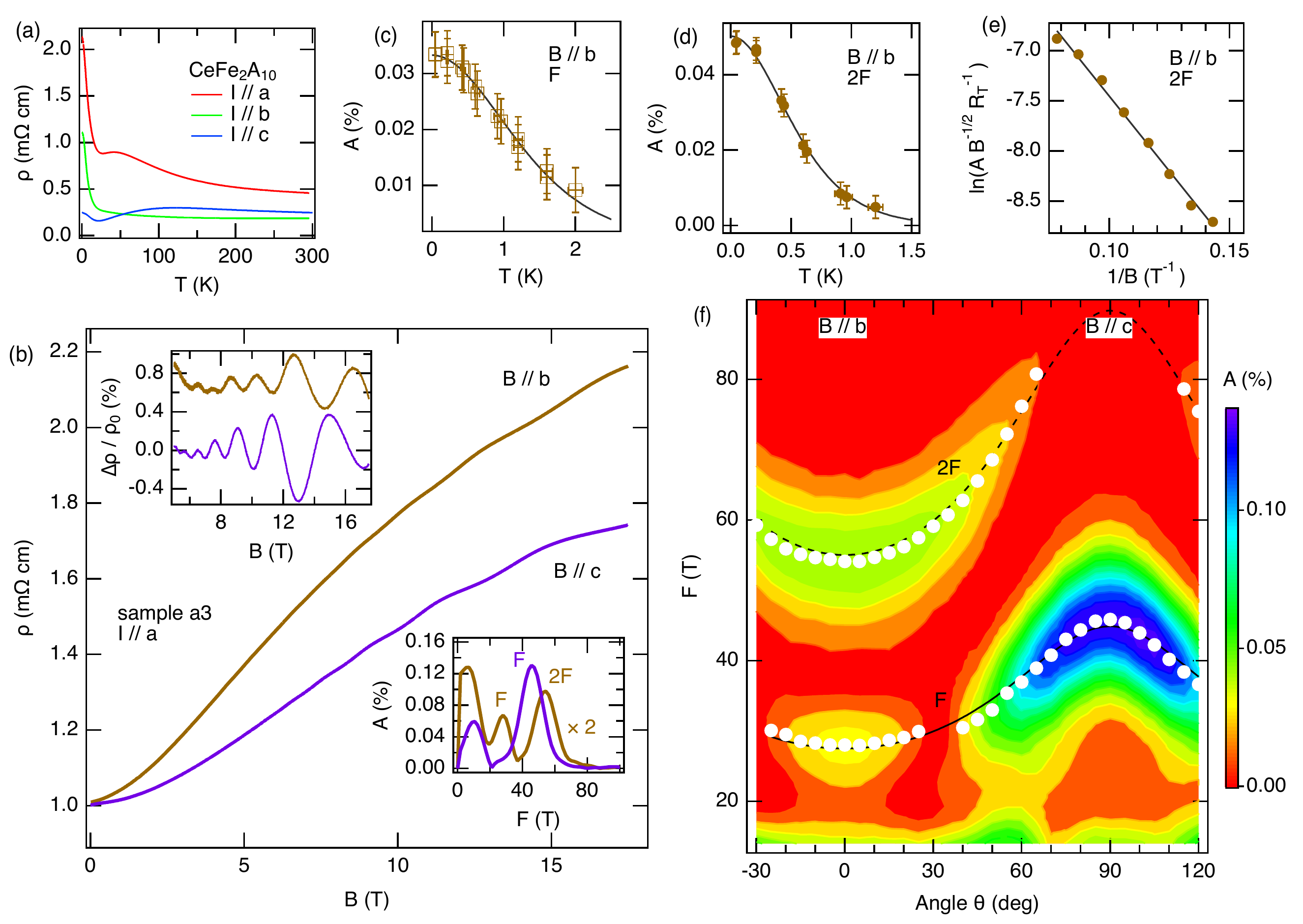}
\caption{\label{Fig2}
Shubnikov--de Haas oscillations in CeFe$_2$Al$_{10}$ at ambient pressure.
(a) Temperature dependence of the resistivity along the three axes of CeFe$_2$Al$_{10}$.
Samples a1, b4, and c3 were used for $I \parallel a$, $b$, and $c$, respectively.
(b) Magnetic-field dependence of the resistivity in sample a3 for $B \parallel b$ and $c$.
The upper inset displays the oscillatory part $\Delta\rho$ normalized by the smooth background $\rho_0$, which was fitted to a third-order polynomial.
The $B \parallel b$ curve is offset for clarity.
The lower inset displays the Fourier transform of $\Delta\rho/\rho_0$ versus $1/B$ in the field range 5--17.5 T.
(c), (d) Temperature dependence of the oscillation amplitude for (c) $F$ = 28 T and (d) $F$ = 54 T for $B \parallel b$.
Solid lines are fits to the temperature reduction factor $R_{T, r}$.
(e) Magnetic-field dependence of the oscillation amplitude of $F$ = 54 T for $B \parallel b$.
The solid line is a fit to the Dingle reduction factor $R_{D, r}$.
(f) Color plot of the Fourier amplitude versus field angle measured from the $b$ axis to $c$.
Measurements were performed every 5$^{\circ}$.
The white circles indicate peaks in the Fourier amplitude spectra.
The solid line is a fit to an ellipsoidal Fermi pocket, while the dashed line is the expected second harmonic.
}
\end{figure*}

\begin{figure*}
\includegraphics[width=16cm]{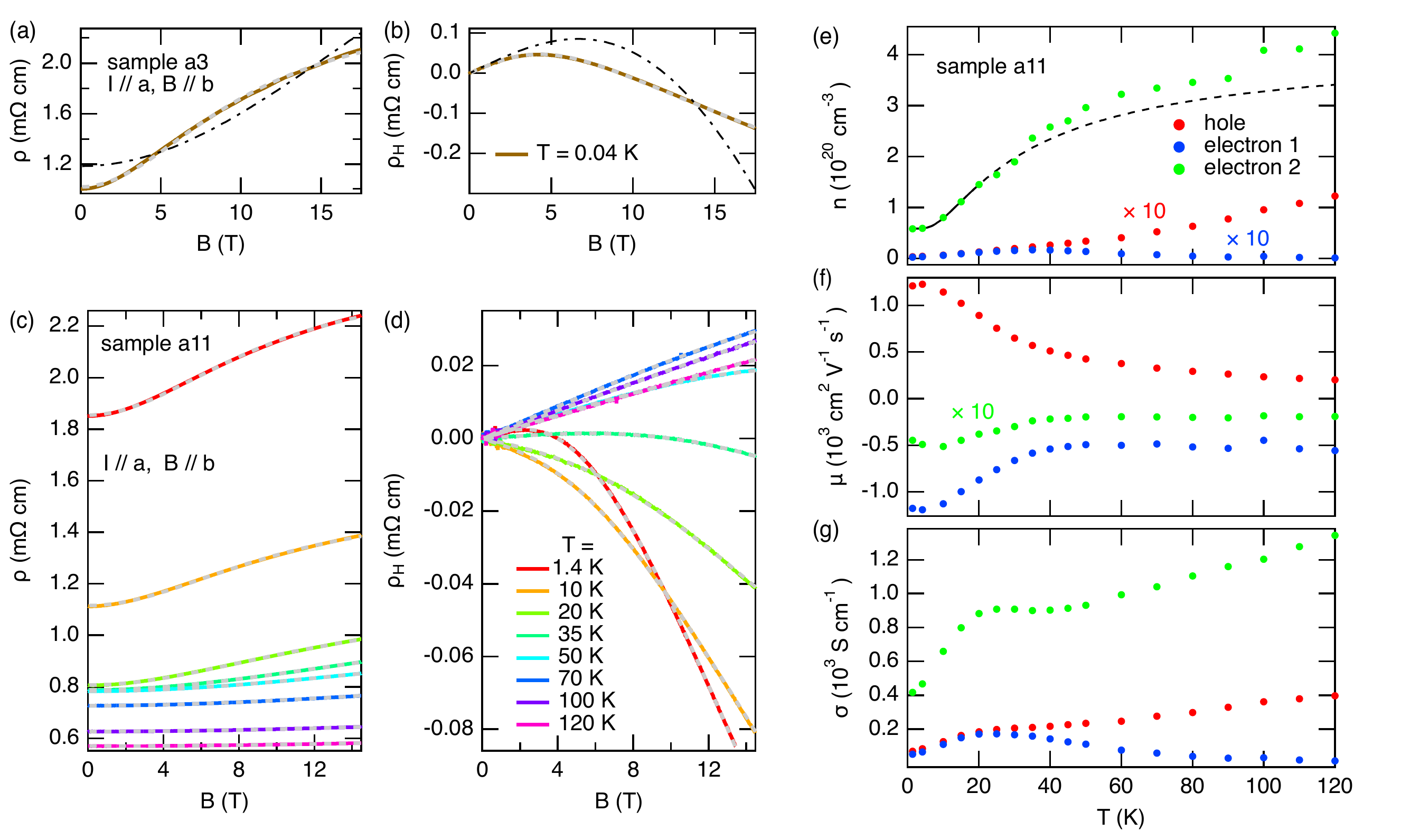}
\caption{\label{Fig3}
Magnetotransport data of CeFe$_2$Al$_{10}$ at ambient pressure.
(a) Resistivity and (b) Hall resistivity of sample a3 at $T$ = 0.04 K.
$I \parallel a$ and $B \parallel b$.
Brown lines represent the experimental data, while
dash-dotted lines are fits to the two-carrier model.
Gray dashed lines are fits to the three-carrier model, which almost completely overlap the brown experimental lines.
(c) Resistivity and (d) Hall resistivity of sample a11 measured at various temperatures (indicated).
Colored lines represent the experimental data, while gray dashed lines are fits to the three-carrier model, which almost completely overlap the colored lines.
(e) Carrier density, (f) mobility, and (g) conductivity determined from the three-carrier fits as a function of temperature.
The black solid (up to 20 K) and dashed (above 20 K) line in (e) exhibits activation-type behavior with an excitation energy $E_{ex}$ = 28 K.
}
\end{figure*}

\begin{figure*}
\includegraphics[width=11cm]{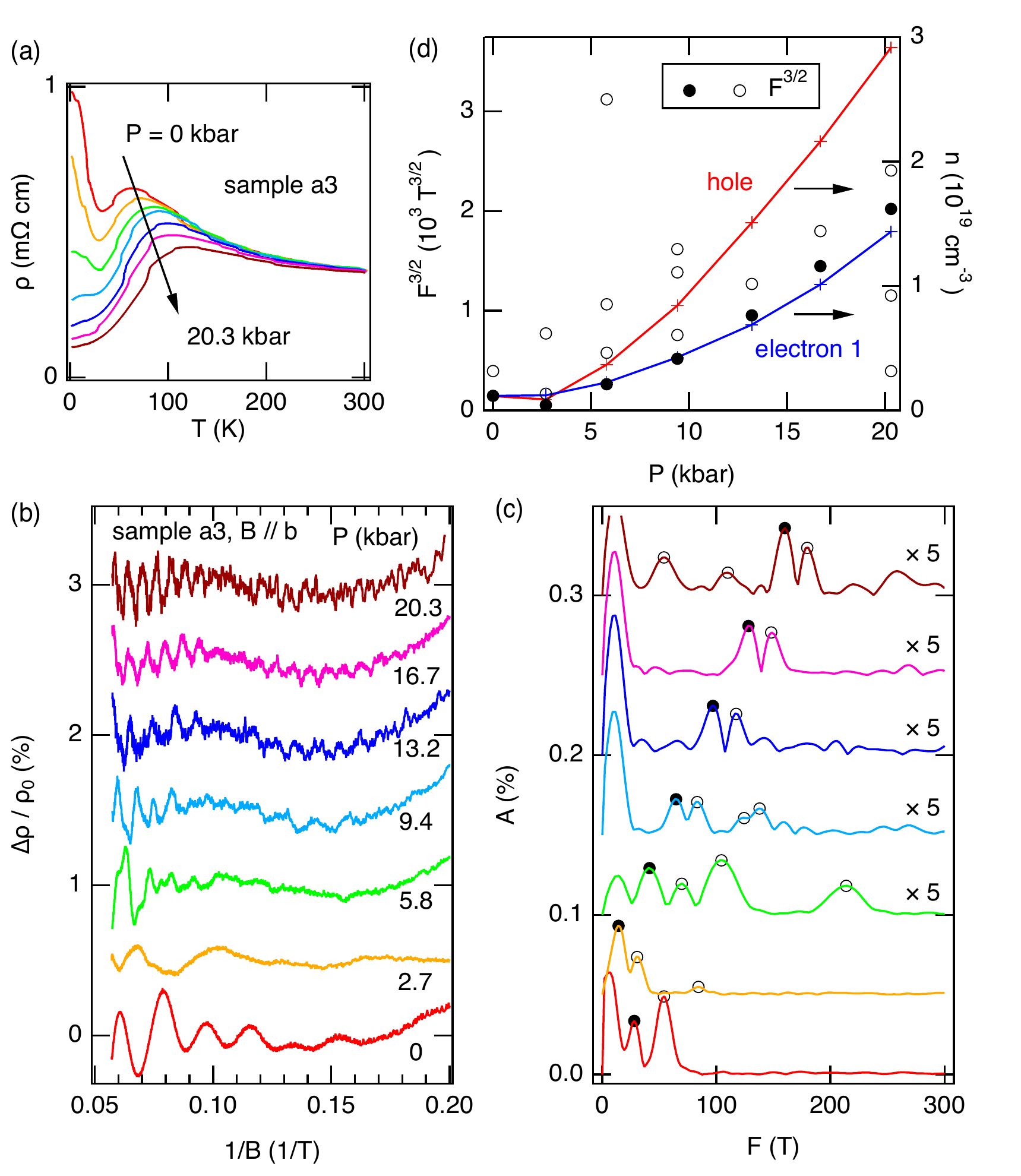}
\caption{\label{Fig4}
Pressure variation of Shubnikov--de Haas oscillations in CeFe$_2$Al$_{10}$.
(a) Temperature dependence of the resistivity of sample a3 at various pressures [indicated by the same color coding as in (b)].
Measurements were performed while a top-loading probe was inserted into a dilution refrigerator.
(b) SdH oscillations of sample a3 for $B \parallel b$ as a function of 1/$B$ at various pressures (indicated).
The curves are offset for clarity.
(c) Corresponding Fourier spectra (offset for clarity).
(d) $F^{3/2}$ of frequency peaks marked in (c) as a function of pressure (solid and open circles).
Red and blue lines indicate the pressure dependence of the minor hole and electron (e1) densities, respectively, determined from the multicarrier analysis of the magnetotransport data.
}
\end{figure*}

\begin{figure*}
\includegraphics[width=11cm]{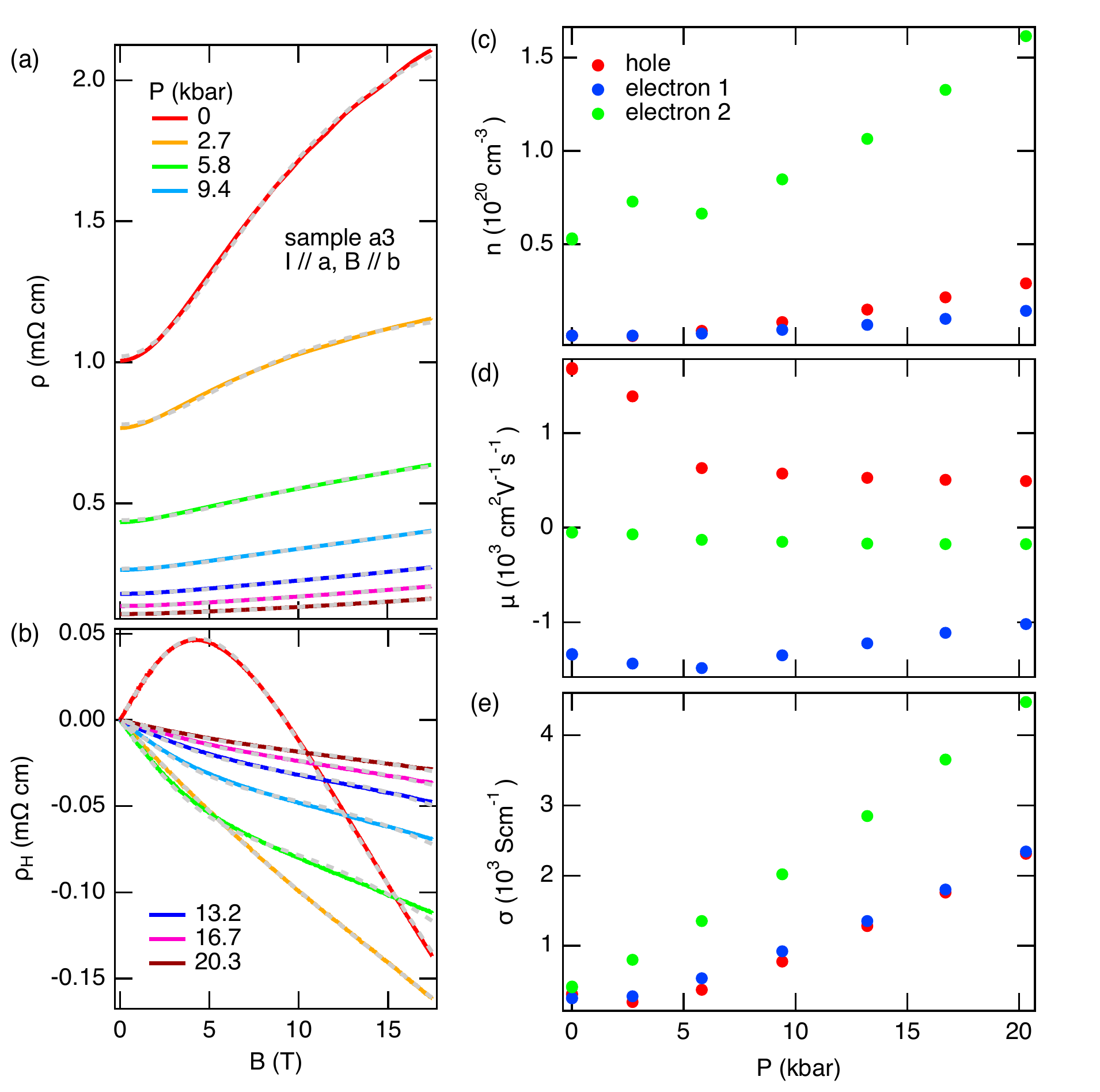}
\caption{\label{Fig5}
Pressure variation of resistivity and Hall resistivity in CeFe$_2$Al$_{10}$.
(a and b) Resistivity (a) and Hall resistivity (b) as a function of magnetic field applied along the $b$ axis at various pressures (indicated).
The temperature was 0.22 K for $P$ = 16.7 kbar and 0.03 or 0.04 K for the other pressures.
(c, d, and e) Carrier density (c),  mobility (d), and conductivity (e) determined from the three-carrier fits as a function of pressure.
}
\end{figure*}

\begin{figure*}
\includegraphics[width=8cm]{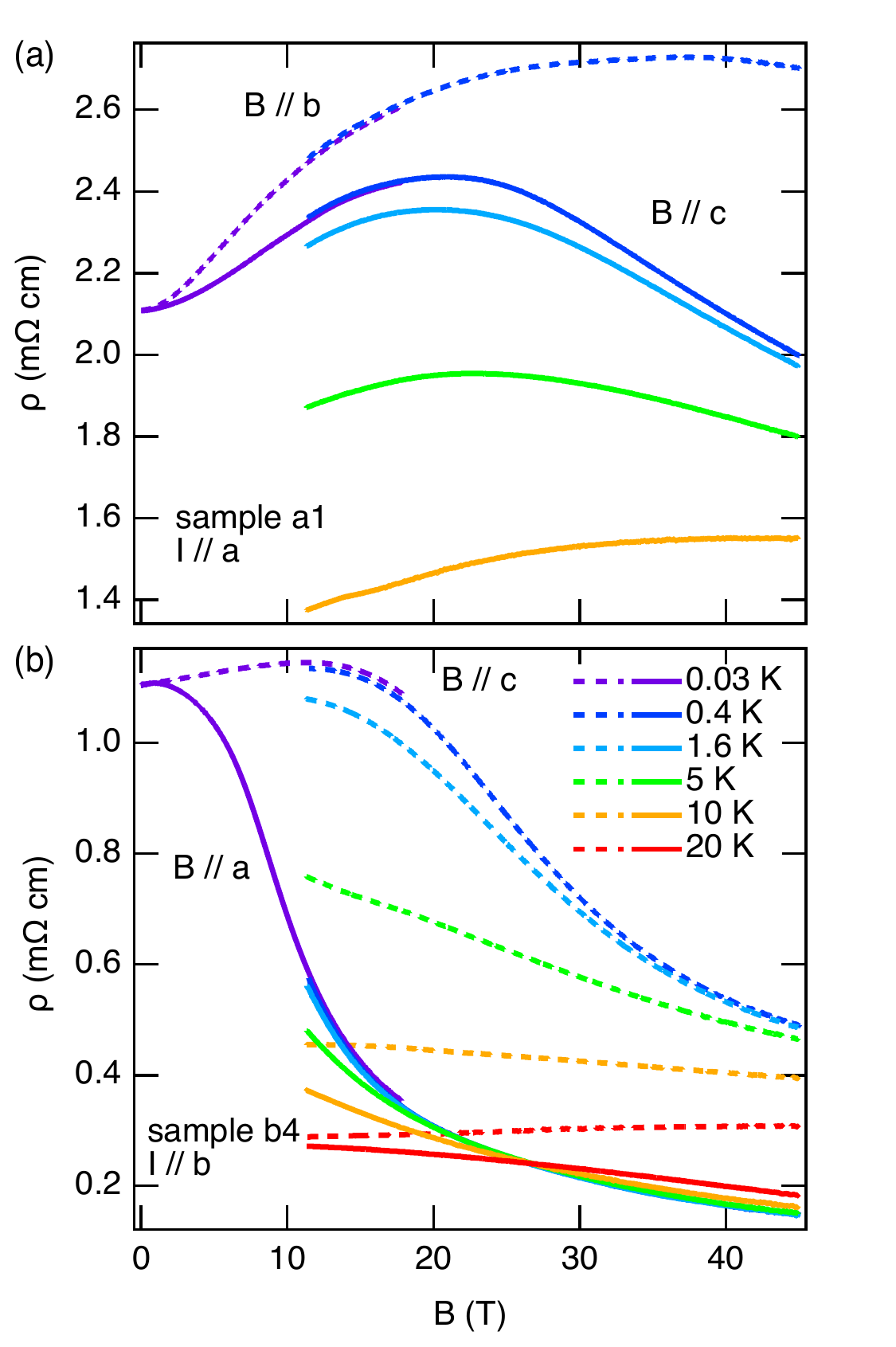}
\caption{\label{Fig6}
Magnetic-field dependence of the resistivity up to 45 T in CeFe$_2$Al$_{10}$.
(a) $a$-axis resistivity as a function of the magnetic field for $B \parallel b$ (dashed lines) and $c$ (solid lines).
(b) $b$-axis resistivity as a function of the magnetic field for $B \parallel c$ (dashed lines) and $a$ (solid lines).
Measurements in a hybrid magnet were performed in the field range 11.5--45 T; measurement temperatures are indicated by the line colors.
Results of measurements at $T$ = 0.03 K in a dilution refrigerator up to 17.5 T are also presented for comparison.
}
\end{figure*}

\begin{figure*}
\includegraphics[width=8cm]{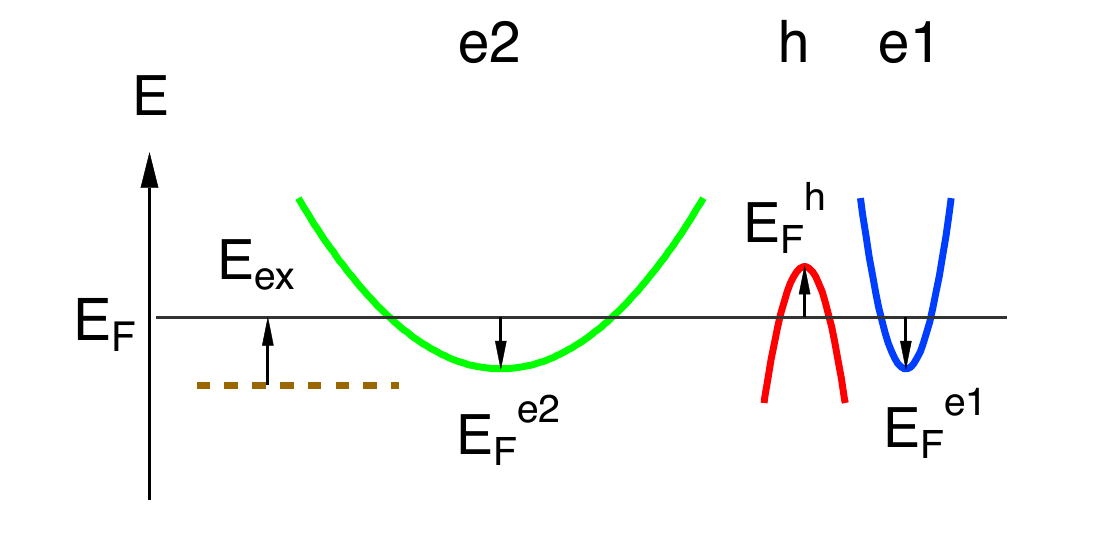}
\caption{\label{Fig7}
Schematic of the electronic structure near the Fermi level deduced from the experimental results.
}
\end{figure*}

\begin{table}
\caption{\label{Tab} Samples.  $l$ is the distance between two voltage contacts, while $w$ and $t$ are sample width and thickness.}
\begin{ruledtabular}
\begin{tabular}{ccccc}
Current direction & Sample & $l \times w \times t$ & $\rho$(0.05 K) & $\rho$(290 K) \\
 & & (mm$^3$) & (m$\Omega$ cm) & (m$\Omega$ cm) \\
 \hline
$I \parallel a$ & a1 & 0.70 $\times$ 0.56 $\times$ 0.22 & 2.1 & 0.46\\
 & a3 & 0.32 $\times$ 0.38 $\times$ 0.18 & 1.0 & 0.38\\
 & a11 & 0.48 $\times$ 0.48 $\times$ 0.22 & 1.9\footnotemark[1] & 0.39\\
 \\
 $I \parallel b$ & b4 & 0.56 $\times$ 0.48 $\times$ 0.16 & 1.1 & 0.19\\
\\
$I \parallel c$ & c3 & 0.68 $\times$ 0.72 $\times$ 0.12 & 0.25 & 0.25\\

\end{tabular}
\end{ruledtabular}
\footnotetext[1]{measured at 1.4 K}
\end{table}

\end{document}